\pdfoutput=1
\documentclass[
 pra, reprint,
 superscriptaddress,
 amsmath,amssymb,
 aps,longbibliography
]{revtex4-2}

\usepackage{graphicx}
\usepackage{amsmath}
\usepackage{xcolor}
\usepackage[bookmarks=false]{hyperref}
\usepackage[normalem]{ulem}
\usepackage{soul}

\newcommand{\ket}[1]{\mathinner{|#1\rangle}}
\begin{document}

\title{Fast single-qubit gates for continuous dynamically decoupled systems}

\author{Michael Senatore}
\email{Michael.Senatore@us.af.mil}
\affiliation{Air Force Research Laboratory, Information Directorate, Rome NY 13441 USA}
\affiliation{Department of Physics, Syracuse University, Syracuse NY 13244-1130 USA}
\author{Daniel L. Campbell}
\email{Daniel.Campbell.22@us.af.mil}
\affiliation{Air Force Research Laboratory, Information Directorate, Rome NY 13441 USA}
\author{James A. Williams}
\affiliation{Air Force Research Laboratory, Information Directorate, Rome NY 13441 USA}
\author{Matthew D. LaHaye}
\email{Matthew.LaHaye@us.af.mil}
\affiliation{Air Force Research Laboratory, Information Directorate, Rome NY 13441 USA}
\affiliation{Department of Physics, Syracuse University, Syracuse NY 13244-1130 USA}
\date{\today}
\begin{abstract}

Environmental noise that couples longitudinally to a quantum system dephases that system and can limit its coherence lifetime. 
Performance using quantum superposition in clocks, information processors, communication networks, and sensors depends on careful state and external field selection to lower sensitivity to longitudinal noise. In many cases time varying external control fields--such as the Hahn echo sequence originally developed for nuclear magnetic resonance applications--can passively correct for longitudinal errors. There also exist continuous versions of passive correction called continuous dynamical decoupling (CDD), or spin-locking depending on context. However, treating quantum systems under CDD as qubits has not been well explored. Here, we develop universal single-qubit gates that are ``fast'' relative to perturbative Rabi gates and applicable to any CDD qubit architecture. We demonstrate single-qubit gates with fidelity $\mathcal{F}=0.9947(1)$ on a frequency tunable CDD transmon superconducting circuit operated where it is strongly sensitive to longitudinal noise, thus establishing this technique as a potentially useful tool for operating qubits in applications requiring high fidelity under non-ideal conditions.
\end{abstract}
\maketitle
\section{Introduction}

\begin{figure*}
\begin{centering}
\includegraphics[width=7in]{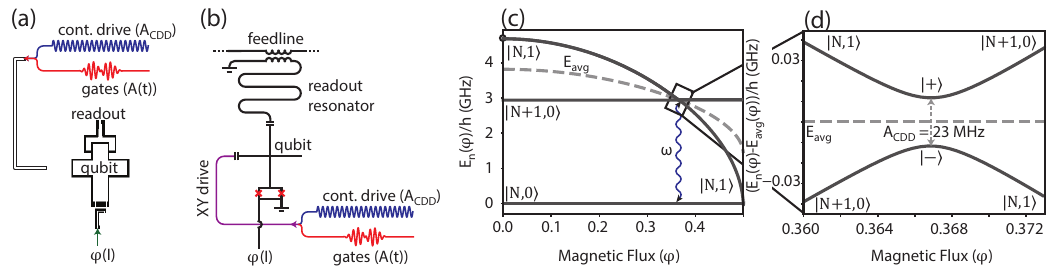}
\caption{
\textbf{(a)} Layout and \textbf{(b)} circuit descriptions of the flux tunable superconducting transmon qubit and the microwave control signals used to create and operate a CDPQ. A transmon is illuminated through a charge bias line by the CDD drive, a continuous wave microwave drive (blue) at frequency $\omega$ and amplitude $A_{CDD}$. Gating pulses $A(t)$ (red) are applied at the same frequency and 90 degrees out of phase with the CDD drive.
\textbf{(c)} In the rotating frame, the CDD drive hybridizes the transmon's excited and ground states. Here, $|N\rangle$ represents the instantaneous number of photons in the waveguide carrying the CDD drive.
\textbf{(d)} Subtracting the average energy of the two hybridized states ($\ket{+}, \ket{-}$) reveals that the relative energy difference between them is first order insensitive to detuning between the transition and $\omega$.
}
\label{fig:fig1}
\end{centering}
\end{figure*}  

Coherence lifetime is a measure of the timescale over which quantum information is preserved in a quantum system. We control a quantum system by making it sensitive to external fields, consequently admitting noise processes that degrade its coherence lifetime.
``Sweet spots'' describe parameter values at which the quantum system becomes first-order insensitive to such an external field, minimizing coherence degradation resulting from noise processes present on that field. 

For any quantum system it is possible to generate protected subspaces that are first-order decoupled from external fields using dynamical decoupling pulse sequences~\cite{viola1998dynamical, viola1999dynamical, viola1999universal} - akin to Hahn Echo and other techniques developed for control and enhancement of coherence in NMR systems~\cite{hahn1950spin, slichter2013principles}. These techniques effectively produce a sweet-spot at any desired external field bias. Increasing the number of decoupling pulses and carefully choosing their interval can improve protection from low frequency noise such as $1/f$~\cite{Siddiqi2021, yoshihara_flux_2014}, which is observed to be ubiquitous across physical systems. The continuum limit of dynamical pulse protection is sometimes called continuous dynamical decoupling (CDD), or ``spin locking,'' which exhibits coherence enhancing properties that have been demonstrated in NMR, NV centers~\cite{avinadav_time-optimal_2014}, superconducting circuits~\cite{gustavsson_driven_2012, gustavsson_suppressing_2016, bylander_dynamical_2011, yan_rotating-frame_2013, yan_flux_2016}, Bose-Einstein condensates~\cite{Trypogeorgos2018}, and other quantum systems~\cite{smirnov_decoherence_2003, abdurakhimov_driven-state_2020}.

Figure~\ref{fig:fig1} illustrates an example of CDD for the case of a superconducting transmon qubit. In this scenario, an oscillating external field applied along the transmon's $xy$~line resonantly drives the transition between its lowest levels (i.e. $|0\rangle\leftrightarrow|1\rangle$), hybridizing the two states. We refer to this external field as the CDD drive.
The splitting of the hybridized states $|\pm\rangle$ is proportional to the amplitude of the applied drive~$A_{CDD}$, equivalent to the Rabi nutation rate during resonant pulsed excitation. The drive generates an artificial sweet-spot [Fig.~\ref{fig:fig1}b-d] in the transition between $|\pm\rangle$ which grants near-dc noise protection: while Fig~\ref{fig:fig1}c-d show protection only as a function of flux, the protection applies to any fields that change the $|0\rangle\leftrightarrow|1\rangle$ transition frequency relative to the CDD drive frequency $\omega$.

While some prior works have used CDD as a noise-spectrometer~\cite{yan_distinguishing_2018, yan_rotating-frame_2013, von_lupke_two-qubit_2020, sung_multi-level_2021, yan_flux_2016}--which only requires initialization of the CDD without the need for gating techniques--other prior works have explored using a pair of CDD states as a high coherence qubit~\cite{awschalom_development_2021, avinadav_time-optimal_2014, Zuk2024, Huang2021}. Due to the small transition between CDD states Rabi gates are prohibitively slow, characterized by a tradeoff between gate fidelity and speed. When the Rabi approximation is violated--that is, when the Rabi rate is comparable to the transition frequency--polar and azimuthal rotations on the Bloch vector evolve nonlinearly with $xy$ pulse amplitude. 

Building on Refs.~\cite{avinadav_time-optimal_2014, campbell_universal_2020, Zhang2021}, the latter two of which demonstrate universal control for small-gapped systems, we adapt Ref.~\cite{campbell_universal_2020}, specifically, to the CDD system and then demonstrate universal gating near the ``speed limit''~\cite{Margolus1998SpeedLimit}. We concretely define the ``speed limit'' in section~\ref{sub:ideal}. In so doing, the number and fidelity of single-qubit operations performable during the coherence lifetime of a CDD system is increased to near its theoretical maximum. This capability allows sequences of time domain gates to be performed quickly and with high fidelity on the wide variety of quantum systems whose coherence can be improved by CDD, see for example Refs.~\cite{miao_universal_2020, abdurakhimov_driven-state_2020, yan_flux_2016}.

In the following, we discuss the general methods for performing universal gates on Continuous Dynamical-decoupling Protected Qubits (CDPQs) generated on two-level superconducting quantum systems and detail how we modify the preparation and gating of the CDPQ in the presence of a third non-computational state, a feature of transmon circuit level structure.
Moreover, we demonstrate that the characteristic noise protection of a CDPQ improves the Hahn echo and Ramsey coherence times by more than a factor of ten for a transmon operated at a flux sensitive point. Finally, we demonstrate universal single qubit gates on a CDPQ, which may have applications towards quantum information processing on superconducting circuits--and many other quantum platforms--in otherwise noisy environments.

\section{Description of the system}\label{system}

The transmon superconducting circuit test device--fabricated by Rigetti Quantum Foundry--is frequency tunable by means of magnetic flux threaded through a symmetric junction superconducting quantum interference device (SQuID). The transmon is a weakly anharmonic oscillator~\cite{koch2007charge} and, in the presence of the time-dependent CDD and gating microwave driving fields, may be approximately described by the Hamiltonian
\begin{align}
    H/\hbar =& \omega_q a^{\dagger} a -(E_C/\hbar) a^{\dagger}a^{\dagger}aa \label{eq:transmonH} \\
    +&A_{CDD} (a +a^{\dagger})\cos{(\omega t)}\nonumber\\
    +&A(t) (a + a^{\dagger})\sin{(\omega t)}\nonumber,\\
    \omega_q(\varphi) =&\omega_0\sqrt{\cos{(\pi\varphi)}}-E_C/\hbar,
\end{align}
where $a^{\dagger}$ and $a$ are the raising and lowering operators for a harmonic oscillator, $\varphi = \Phi/\Phi_0$ is a dimensionless flux parameter where $\Phi_0$ is the magnetic flux quantum, $\omega_0/2\pi-E_C/h = 4.64$~GHz is the transition frequency of the transmon at its flux upper sweet spot (USS), and $E_C/h = 137$~MHz is the anharmonicity. The transmon is Purcell limited to $T_{1}\sim 20\,\mu s$ by its readout resonator $5.86$~GHz at its USS. At the USS the Hahn echo coherence $T_{2E} \sim 40\,\mu s$ is $T_1$ limited. For the purposes of exploring noise mitigation with CDD, the transmon is flux-tuned to $2.9$~GHz. At this flux and frequency, the $T_{1}$ increases to $66(12)\,\mu s$ because the transmon is further detuned from its readout resonator, while the coherence becomes limited by flux-noise in SQUID such that the Ramsey $T_{2R} = 1.29(4)\,\mu s$ and Hahn echo $T_{2E} = 4.4(2)\,\mu s$.

Once initialized, the CDD drive with amplitude $A_{CDD}$ remains always on at frequency $\omega$, providing protection for the CDPQ when $\omega \approx \omega_q$. A substantial contribution of this work is showing that gating pulses $A(t)$ applied at the same frequency and 90~degrees out of phase with $A_{CDD}$ can gate the CDPQ with high fidelity.

\begin{figure}
\includegraphics[width=3.3in]{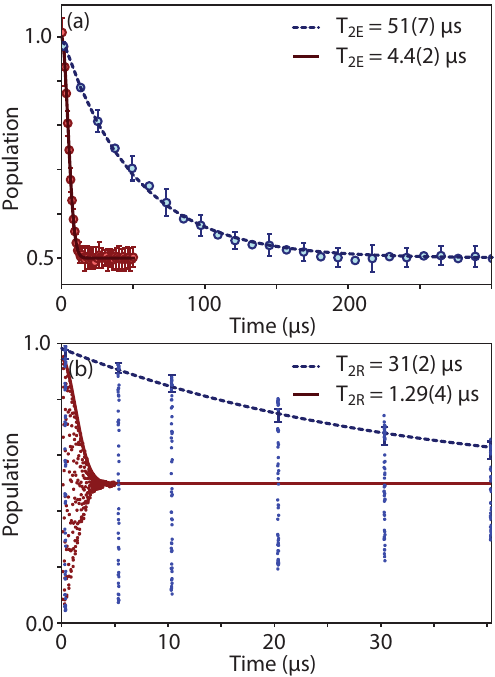}
\caption{
Comparison of coherence of bare tunable transmon (red markers) and CDPQ (blue markers) at $\varphi = 0.367$. The solid red (dashed blue) traces represent a Gaussian (exponential) fits to the bare transmon (CDPQ) data. The error bars correspond to the standard deviation of 36 repetitions of the measurement. Panel \textbf{(a)} compares Hahn echo coherence while \textbf{(b)} compares Ramsey coherence. The CDPQ Ramsey sequence was sampled over $100\text{ ns}$ starting at $0.3$, $5.3$, $10.3$, $20.3$, $30.3$, and $40.3\text{ }\mu\text{s}$
}
\label{fig:fig2}
\end{figure}  

\subsection{Noise protection}

We apply $A_{CDD}$ to the transmon qubit via a local drive line as shown in Fig.~\ref{fig:fig1}a-b. The driving field hybridizes the bare transmon $\ket{0}\leftrightarrow \ket{1}$ transition, opening an avoided level crossing with gap $A_{CDD}$ as shown in Fig.~\ref{fig:fig1}c-d. The separation between the hybridized states, represented in Fig.~\ref{fig:fig1}d, is first-order insensitive to fluctuations in $\omega - \omega_q(\varphi)$, translating to an effective insensitivity to $\varphi$, near the $\varphi$ that achieves minimum separation (the sweet spot). At drive amplitude $A_{CCD} = 2\pi\times 23$~MHz the CDPQ has a lifetime of $T_1 = 48(10)\,\mu s$ and the Ramsey and Hahn echo coherence times increase to $T_{2R} = 31(2)\,\mu s$ and $T_{2E}=51(7)\,\mu s$, respectively, as shown in Fig.~\ref{fig:fig2}.
Dephasing of the CDPQ states scales linearly with noise on $A_{CDD}$ and so this protocol trades sensitivity to frequency for sensitivity to amplitude on the drive-tone. For most precision microwave sources, the power spectral density of amplitude noise is lower than that of phase noise, and so this would generally be an advantage, especially when the qubit's transition frequency is exposed to a strong source of $1/f$ noise as is the case with a detuned flux-tunable transmon.

\subsection{Initialization}

\begin{figure}
    \centering
    \includegraphics[width = 3.3in]{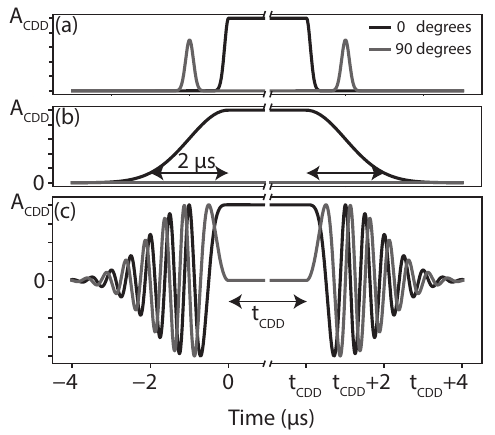}
    \caption{\textbf{(a)} State preparation using standalone Rabi pulses. The carrier phase of the gray pre- and post-pulses is set 90 degrees out of phase from the CDD drive to map between the bare transmon and CDPQ bases~\cite{yan_rotating-frame_2013}. \textbf{(b)} Adiabatic state preparation used in data collection and suitable for use with multi-level systems. The CDD drive is ramped on and off with $2\,\mu s$ long half-Gaussian waveform. \textbf{(c)} Alternative adiabatic state preparation approach. The CDD drive is increased while off-resonant, then chirped over $1\,\mu s$ into resonance.}
    \label{fig:Init}
\end{figure}

Pulsed spin locking preparation, shown in Fig.~\ref{fig:Init}a, can effectively prepare a CDPQ on quantum systems with high anharmonicity~\cite{yan_flux_2016,yan_distinguishing_2018,yan_rotating-frame_2013,gustavsson_suppressing_2016, gustavsson_driven_2012}. The pre- and post-pulses in this protocol map between the bare qubit basis and the driven eigenbasis of the CDPQ. In transmons, the CDPQ eigenbasis includes non-negligible contributions from $|2\rangle$, making the CDPQ eigenbasis difficult to prepare with pre- and post-pulses that only perform rotations within the $|0\rangle$ and $|1\rangle$ bare transmon subspace. For this reason, the initialization procedure shown in Fig.~\ref{fig:Init}a cannot prepare an eigenbasis of the CDPQ on a transmon. An easy way to check for preparation of a non-eigenbasis is to perform a single-CDPQ gate at time $t_0$ for $0<t_0<t_{CDD}$: oscillations in the bare transmon state population will appear as $t_{CDD}$ is varied. We see these oscillations using the Fig.~\ref{fig:Init}a protocol and further quantify this initialization mismatch in section~\ref{sub:anharmonicityleakage}.

Hence, we use the Gaussian envelope with a flat top to adiabatically prepare and then hold the CDPQ, as shown in Fig.~\ref{fig:Init}b, which automatically performs the required mapping to and from the CDPQ eigenbasis and does not have the initialization issue described above. This adiabatic state preparation protocol relies on $A_{CDD}$ being a significant fraction of the anharmonicity. We ramp $A_{CDD}$ \textit{off-resonant} from the transmon $|0\rangle\leftrightarrow|1\rangle$ transition. As the amplitude increases, interaction with the $|1\rangle\leftrightarrow|2\rangle$ transition shifts the $|0\rangle\leftrightarrow|1\rangle$ transition into resonance with the drive. Adiabatically chirping $\omega$, as shown in Fig.~\ref{fig:Init}c, theoretically allows for faster CDPQ initialization. We tested this approach and found it effective, however, all data in this manuscript uses the approach in Fig.~\ref{fig:Init}b because it requires tuning fewer parameters. A well optimized adiabatic initialization using frequency chirping or flux sweeping can be performed in time $\sim 4\pi/A_{CDD}$~\cite{MartinisCZ, DiCarlo2009}. When so optimized, this duration is fairly short and would be compatible with high process-fidelity operation.

\section{Universal single-qubit gates}

\subsection{Ideal two-level system}
\label{sub:ideal}

\begin{figure}
\begin{centering}
\includegraphics[width=3.3in]{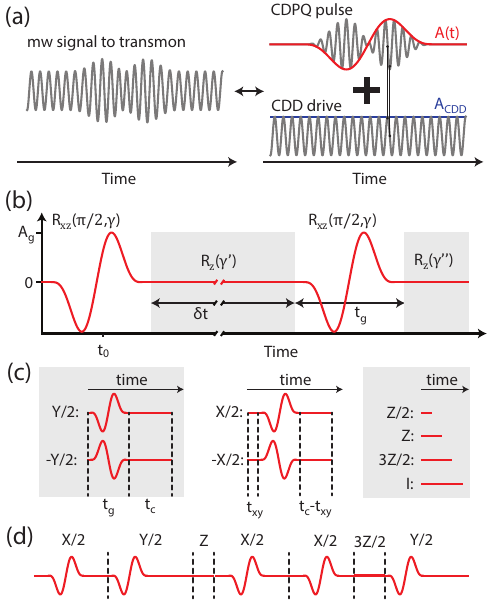}
\caption{
\textbf{(a)} The microwave driving fields sent to the transmon can be decomposed into two components. The first is the CDD drive that generates the CDPQ. The second consists of pulses that gate the CDPQ. These two drives have the same frequency but are 90 degrees out of phase as shown by the vertical black lines. \textbf{(b)} We implement universal gates on a CDPQ by varying the wait times between two fixed amplitude pulses. \textbf{(c)} Combinations of pulses and wait times can be composed into traditional single-qubit gates. \textbf{(d)} 
}
\label{fig:gates}
\end{centering}
\end{figure}

In this section we consider the case of a high anharmonicity system where the CDPQ can be approximated as an ideal two-level system as shown in Fig.~\ref{fig:fig1}d. The more complex case of a low anharmonicity system is treated in the subsequent section.

The CDD drive $A_{CDD}$ and the gate pulses $A(t)$ together comprise the total driving field whose phase and magnitude are depicted in Fig.~\ref{fig:gates}a. We write an effective CDPQ Hamiltonian in the basis defined by the $A_{CDD}$ driving field and in the frame of reference rotating with $\omega$. In this basis $A(t)$ and flux control $\Delta(t) = \omega - \omega_q(\varphi(t))$ act as orthogonal transverse fields
\begin{align}
    H/\hbar =& A_{CDD} \sigma_{z}/2+A(t)\sigma_{x}/2+\Delta(t) \sigma_{y}/2\label{eq:controlideal}
\end{align}
where $\sigma_{x,y,z}$ are the Pauli matrices.

Universal gates on the single-qubit Bloch sphere can be efficiently performed with sequences of arbitrary $R_z(\phi)=\text{exp}{(-i\phi\sigma_{z}/2)}$ rotations interspersed with a single $R_{xz}(\theta, \phi)=\text{exp}{(-i\theta\sigma_{x}/2-i\phi\sigma_z/2)}$ gate where $\theta$ and $\phi$ correspond to polar and azimuthal rotations on the Bloch sphere, respectively. For $R_{xz}$ rotations we require that $\theta = \pi/2$ and note that there will be an initially unknown azimuthal rotation $\phi = \gamma$. An arbitrary universal gate takes the form (also see the supplemental materials of Ref.~\cite{campbell_universal_2020})
\begin{align}
    U(\gamma', \gamma'') = R_{xz}(\pi/2, \gamma)R_z(\gamma')R_{xz}(\pi/2, \gamma)R_{z}(\gamma'')\label{eq:U}.
\end{align}

We perform $R_z(\phi)$ rotations by varying wait times between $R_{xz}(\pi/2, \gamma)$ pulses, utilizing the linear evolution of the phase according to $\partial \phi/\partial t = A_{CDD}$. $R_{xz}(\pi/2, \gamma)$ rotations are performed by pulsing $A(t)$ with a cosine and sine multiplied together with peak amplitude $A_g$ (Fig.~\ref{fig:gates}b) over the domain ${|t-t_0|\leq t_g/2}$
\begin{align}
    A(t) =& \frac{A_g}{1.3} (1+\cos{2\pi (t-t_0)/t_g)})\sin{(2\pi (t-t_0)/t_g)},
\end{align}
transforming the bare qubit states into equal superposition states $\ket{\pm} = (\ket{0}\pm \ket{1})/\sqrt{2}$.
The interplay between $A_g$ and $t_g$ is shown in Fig.~\ref{fig:multilevel}c-d.
$A_{CDD}$ sets a natural timescale for gating. $R_{xz}(\pi/2, \gamma)$ pulse duration $t_g \sim 2\pi/A_{CDD}$ and amplitude $A_g \sim A_{CDD}/2$ approaches this speed limit at the cost of making the pulses non-perturbative (non-Rabi). The absolute ``speed limit'' for a choice of $A_{CDD}$ is the lowest $t_g$ along any 50:50 superposition contour in Fig.~\ref{fig:multilevel}c-d, however, this parameterization becomes insensitive to $A_g$ as a tuning parameter and can lead to high leakage in transmons. We note that modulating $\Delta(\varphi(t))$ rather than $A(t)$, which may be accomplished by flux tuning or frequency chirping $A_{CDD}$, performs a $R_{yz}$ rotation and can replace $R_{xz}$ in the above protocol.

We show in Fig.~\ref{fig:gates}b a pulse sequence comprised of two such $R_{xz}(\pi/2, \gamma)$ rotations, each with a total duration $t_g\sim 2\pi/A_{CDD}$, separated by a variable delay $\delta t$ that performs $R_z(\gamma' = A_{CDD}\delta t)$. While this sequence can prepare any polar angle rotation of the Bloch vector by varying $\delta t$, following the whole sequence with a second $R_z(\gamma'')$ may prepare any arbitrary rotation of the Bloch vector, as claimed in Eq.~\ref{eq:U} and shown in Fig.~\ref{fig:gates}b.

As shown in Fig.~\ref{fig:gates}c we may create the single-qubit $Y/2$ gate by concatenating $R_{xz}(\pi/2,\gamma)$ with $R_z(-\gamma)$. Negative phases may be realized by giving the time delay the form $t_c = 2\pi/A_{CDD}-\gamma/A_{CDD}$. The $X/2$ gate adds $Z/4 = R_z(\pi/2A_{CDD})$ to the beginning of the $Y/2$ and subtracts it from the end. Unlike in Rabi-based control schemes, gates are arranged end-to-end no time separation as shown in Fig.~\ref{fig:gates}d. Note that while $\pm X/2$ and $\pm Y/2$ gates are uniform in duration, $Z$ gates are not.

\subsection{Multi-level system}

\begin{figure}
\begin{centering}
\includegraphics[width=3.3in]{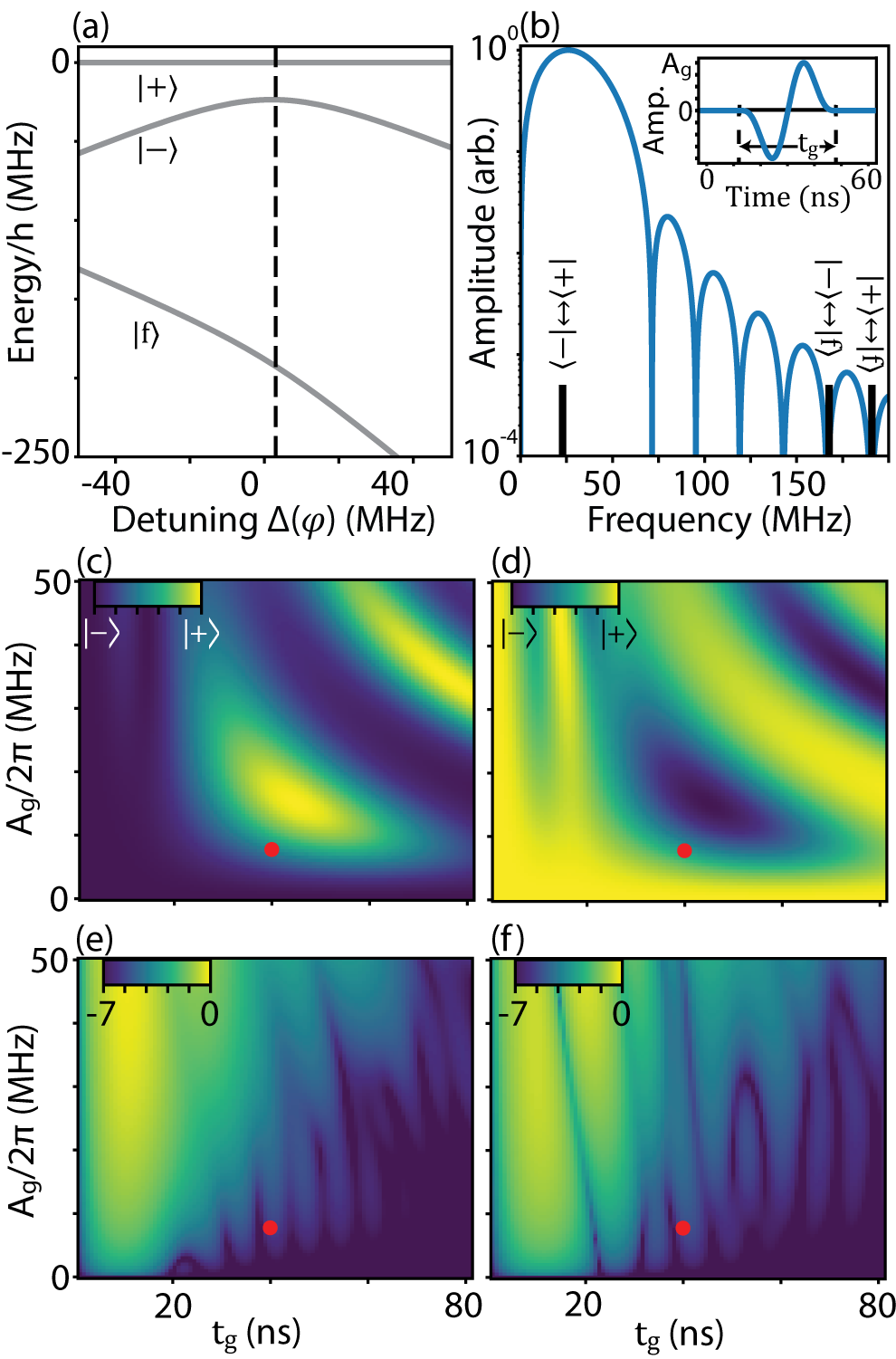}
\caption{
\textbf{(a)} Level structure in the driven frame showing the first three eigenenergies (solid gray lines) as a function of the drive's detuning $\Delta(\varphi) = \omega - \omega_q(\varphi)$.
\textbf{(b)} A fast Fourier transform (FFT) of the CDPQ gate-pulse temporal profile (shown in the inset), where $\tau = 40\text{ ns}$ was the pulse duration used for data collection. The overlap of transitions between eigenstates with nodes (anti-nodes) of the FFT can predict the absence (prevalence) of leakage to state $\ket{f}$. This picture is approximate because the eigenenergies shift as a function of pulse amplitude $A_g$ and is most accurate when the gate-pulse has a perturbative effect on eigenstate $\ket{f}$.
\textbf{(c, d, e, f)} Simulation of transmon state population as a function of $R_{\theta,\phi}$ pulse width $t_g$ and pulse amplitude $A_g$. The red dots mark the parameters of $A_g$ and $t_g$ used in this work and were sufficient to eliminate leakage as a dominant source of infidelity. The left panels \textbf{(c, e)} correspond to initial state $\ket{-}$ while the right panels \textbf{(d, f)} correspond to initial state $\ket{+}$. The upper panels \textbf{(c, d)} show the CDPQ state after the pulse. The lower panels \textbf{(e, f)} show the fractional log population $P_f$ of $\ket{f}\sim \ket{2}$. The colorbar for panels \textbf{(e, f)} shows the exponent $x$ such that $P_f = 10^{x}$.
}
\label{fig:multilevel}
\end{centering}
\end{figure}

Compared to the case of an ideal two-level system that was analyzed in the previous section, creating a CDPQ on a transmon superconducting qubit introduces additional complexities owing to the CDD drive and gate pulses coupling to additional, non-computational states.

We analyze the full transmon level structure (Eq.~\ref{eq:transmonH}) in the rotating frame of the drive and apply the rotating wave approximation to trim the counterrotating terms. Deviations from the ideal two-level scenario may be understood by including only the first non-computational transmon state $\ket{f}$ as shown in Fig.~\ref{fig:multilevel}a and represented by the truncated Hamiltonian,
\begin{align}
    &H(t)/\hbar =\label{eq:tranmontruncated}\\
    &\begin{pmatrix}\Delta & A_{CDD}-iA & 0\\
    A_{CDD}+iA & -\Delta & \sqrt{2}(A_{CDD}-iA)\\
    0 & \sqrt{2}(A_{CDD}+iA) & -2E_C/\hbar-3\Delta\end{pmatrix}/2.\nonumber
\end{align}
Note that we dropped the time dependence from $\Delta(t)$ and $A(t)$ for notational simplicity.

A desirable form for the Hamiltonian would reveal the interplay between the control degrees of freedom and the driven eigenstates for a three state system, much as Eq.~\ref{eq:controlideal} did for the ideal two level system. We may obtain an approximate eigenbasis transformation $\check{T}$ for Eq.~\ref{eq:tranmontruncated} in the limit where $\beta = \hbar A_{CDD}/E_C < 0.2$ and temporarily setting $\Delta(t)$ and $A(t)$ to zero for all times.
\begin{align}
    \check{T} \sim \begin{pmatrix} \frac{1}{\sqrt{2}} & -\frac{1}{\sqrt{2}} & \sqrt{2}\beta^2 \\
    -\frac{1}{\sqrt{2}} & -\frac{1}{\sqrt{2}} & -\sqrt{2}\beta\\
   -\beta & -\beta & 1\end{pmatrix}\label{eq:CDPQtransform}
\end{align}
where we neglect the contribution of $\beta$ to the normalization for notational clarity. The CDPQ eigenstates can now be expressed as a combination of the lowest three transmon states ``dressed'' by $A_{CDD}$.
\begin{align}
    \begin{pmatrix}\ket{-} \\
    \ket{+}\\
   \ket{f}\end{pmatrix}=
       \check{T}
       \begin{pmatrix}\ket{0} \\
    \ket{1}\\
   \ket{2}\end{pmatrix}\label{eq:CDPQtransform}.
\end{align}

Transforming Eq.~\ref{eq:tranmontruncated} so that $H' = \check{T}H\check{T}^{-1}$ with $\Delta(t)$ and $A(t)$ reintroduced then reveals
\begin{align}
    H'(t)/\hbar \sim \begin{pmatrix}A_{CDD} & -iA-\Delta & iA+\beta\Delta\\
    iA-\Delta & -A_{CDD} & iA+\beta\Delta\\
    -iA+\beta\Delta & -iA+\beta\Delta & -2E_C/\hbar-3\Delta\end{pmatrix}/2\label{eq:CDPQdiag}.
\end{align}
We see that either time-dependent driving field, $A(t)$ or $\Delta(t)$, can induce transitions between $\ket{+},\,\ket{-}$ and $\ket{f}$.

\subsection{Role of anharmonicity in initialization and leakage}
\label{sub:anharmonicityleakage}

Using Eq.~\ref{eq:CDPQdiag}, we may quantify aspects of our discussion of initialization from section~\ref{system}. In transmons, small anharmonicity and the desire for fast gate speeds (favoring larger $A_{CDD}$) will typically result in a choice of $A_{CDD}$ such that $0.01 < \beta < 0.2$. After the first pulse in the Fig.~\ref{fig:Init}a initialization protocol there is a $\beta$ component orthogonal to the CDPQ basis, constituting a substantial preparation error for our choice of $\beta = 1/7$.

Bare transmon qubits exhibit energy relaxation, often to the electronic ground state $\ket{0}$. While the same process will occur to the bare transmon states in a CDPQ, $\ket{0}$ is distributed principally amongst $\ket{+}$ and $\ket{-}$ with a very small fraction in $\ket{f}$. Therefore, in the steady state the CDPQ relaxes into a $50:50$ incoherent mixture of $\ket{+}$ and $\ket{-}$. The $\beta^4$ fractional ground state $\ket{0}$ population in $\ket{f}$ should result in $0.01\%$ or less steady state population in a non-computational state of the driven system. We did not observe any population in $\ket{f}$ during single shot measurements after randomized benchmarking.

\subsection{Pulse tuning procedure}

Having previously measured the transmon anharmonicity, we select a $A_{CDD}$ and $t_g$ with minimal leakage into the non-computational state $|f\rangle$ from either CDPQ state $|+\rangle$ or $|-\rangle$, simulated in Fig.~\ref{fig:multilevel}e-f. Matching the transitions $|+\rangle \leftrightarrow |f\rangle$ and $|-\rangle \leftrightarrow |f\rangle$ to the nodes of the Fourier transform of the pulse envelope of $A_g$ approximately accomplishes this goal in the limit where $\{A_g,\,A_{CDD}\}\ll E_C/\hbar$. This is roughly equivalent to making $E_C/\hbar A_{CDD} = 7$ an integer ratio.

Using the simulated value for $t_g$, we implement a sequence of two back-to-back $X/2$ gates, defined in Figure~\ref{fig:gates}c, initially guessing at $A_g$ and $t_c$. Sweeping $A_g$ and $t_c$, the sequence implements a Ramsey that oscillates at rate $\sqrt{A_{CDD}^2+\Delta^2(\varphi)}$ with $t_c$. This rough scan checks the value of $A_{CDD}$ and sets $\Delta(\varphi) = 0$. If $A_{CDD}$ meets the criteria above, we then select the $A_g$ that provides the maximal contrast with scanned $t_c$ and the minimum value for $t_c$ that flips the CDPQ state.

We optimize $A_g$ and $t_c$ using trains of $X/2$ gates, iteratively switching between scanning $A_g$ and $t_c$ for successively more gates in the train to obtain additional precision, see also the supplement of Ref.~\cite{ campbell_universal_2020}. Knowledge of $A_{CDD}$ is all that is required to specify $Z$ gates, where $\phi$ evolves with time at rate $A_{CDD}$, and $t_{xy} = \pi/2A_{CDD}$.

\section{Results}

\begin{figure}
    \centering
    \includegraphics[width = 3.3in]{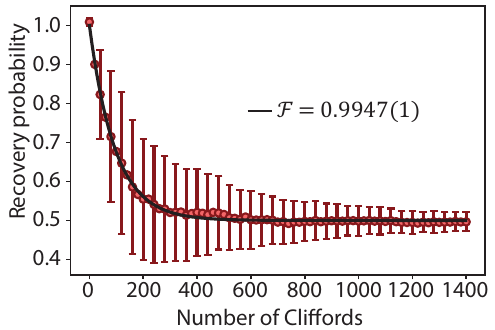}
    \caption{Recovery state population as a function of consecutive single-qubit Cliffords performed on the tunable transmon. Error bars are the standard deviation over 354 different random sequences.}
    \label{fig:FidelityFit}
\end{figure}

After initializing the CDPQ with $A_{CDD}/2\pi = 23\text{ MHz}$, and tuning up the gates with $A_{g}/2\pi = 29.12\text{ MHz}$, we measured the average Clifford fidelity via randomized benchmarking~\cite{Magesan2011,Magesan2012,Gaebler2012,Knill2008,Ryan2009}. The specific randomized benchmarking sequence used for the CDPQ is described in the supplement of Ref.~\cite{campbell_universal_2020}. With Clifford gate-durations of $\sim 93$~ns, we measure an average Clifford gate fidelity of $0.9947(1)$ (total error $5.6\times 10^{-3}$) as shown in Figure~\ref{fig:FidelityFit}. Padding the gate time to $\sim 132$~ns does not reduce the Clifford fidelity, implying that the errors are primarily coherent in origin. CDPQ gate coherence contributes $1.2\times 10^{-3}$ to the Clifford error. We estimate from simulations that leakage to the second excited state introduces error at the $3\times 10^{-4}$ level. We attribute the remaining error $4.1\times 10^{-3}$ to slow changes to the static value of $\Delta$ from flux drift for which we did not compensate. 

The CDPQ plausibly outperforms the undriven transmon. At $\varphi = 0.367$ the undriven transmon is dominated by incoherent errors. With a Hahn Echo lifetime of $4.4\text{ }\mu\text{s}$, Clifford gates on the undriven transmon would need to have an average separation of less than $45\text{ ns}$ to achieve a similar Clifford fidelity to the CDPQ. Moreover, a larger CDPQ gap could provide better protection from flux drift. It would also be possible to periodically measure and correct for flux offsets: as shown in Fig.~\ref{fig:fig2} the instantaneous coherence of the CDPQ is quite good.

For comparison, we also prepared a CDPQ on a fixed frequency transmon, fabricated by MIT Lincoln Laboratory, where slow drifts in $\Delta$ are not present, and obtained a Clifford fidelity of $0.9985$. The bare transmon boasted an average $T_1$ of $112~\mu s$ and an average Hahn echo $T_{2E}$ of $215~\mu s$, which leaves little room for $T_{2E}$ improvement. On this device the CDPQ had a $23$~MHz gap with a $T_{1}$ of $147~\mu s$ and a Hahn echo $T_{2E}$ of $54~\mu s$. This CDPQ $T_{2E}$ is worse than the bare transmon $T_1$ limiting case $T_{2E} \leq 4T_1/3$ predicted in supplementary materials of Ref.~\cite{yan_rotating-frame_2013} and associated references. This sample did not have a local drive line and hence continuous driving the transmon through the readout feedline may have introduced heating that reduced the CDPQ $T_{2E}$.

\section{Conclusions}

We have demonstrated a fast, high-fidelity, approach for performing universal single qubit gates on a CDPQ. The CDPQ under test was generated by driving the lowest transition of a superconducting transmon qubit and showed characteristic protection from frequency noise. The randomized benchmarking Clifford fidelity was consistent with CDPQ frequency noise protection while the transmon was detuned $2\text{ GHz}$ from its frequency sweet spot. The techniques outlined in this paper should be applicable to any physical modality of closed quantum system where a two state transition may be isolated from other transitions and hybridized by a sinusoidal drive.

The CDPQ trades the noise sensitivity to its ambient bath for the amplitude of an external driving field, a field that is often less noisy than the original bath. CDPQs may be an interesting avenue for high coherence operation of quantum systems in the presence of noise they might otherwise be very sensitive to, such as flux noise or charge noise which would typically result in very poor coherence. The low anharmonicity of transmons complicates both initialization and setting large values for $A_{CDD}$ due to state shifting as a function of $A_{CDD}$. Other high coherence superconducting qubit modalities such as C-shunted flux~\cite{yan_flux_2016-1} and fluxonium~\cite{Manucharyan2009} have much larger anharmonicities. Large values of $A_{CDD}$ provide more noise protection and enable faster gates on a CDPQ.

Development of a quantum information processor using CDPQs will require the incorporation of a fast single-qubit gating scheme, such as described in this work, with two-CDPQ gates. Remarkably, there are several experimental demonstrations of two-CDPQ gates~\cite{campbell_universal_2020, Lu2022, Guo2018}. 
From the frequency insensitivity of the CDPQ, we can predict that coupled CDPQs at different frequencies would have suppressed undesireable $\sigma_z\sigma_z$ crosstalk. A viable CDPQ architecture could allow for the elimination of tunable couplers from superconducting circuit quantum processors and add additional flexibility to processor design and to level structure.

\begin{acknowledgments}
Approved for Public Release; Distribution Unlimited: PA\# AFRL-2024-6740. Any opinions, findings, and conclusions or recommendations expressed in this article are those of the authors and do not necessarily reflect the views of the Air Force Research Laboratory (AFRL). The authors would like to thank Yuvraj Mohan and the Rigetti Quantum Foundry Services team for design support and device fabrication of the flux tunable transmon studied in this work. The authors also gratefully acknowledge support from MIT Lincoln Laboratory and IARPA LogiQ program for the single-junction  transmon used in this study and also for the traveling wave parametric amplifiers (TWPAs)) used in the measurement chains for this work. 
\end{acknowledgments}
\appendix
\section{Randomized benchmarking}

\begin{table}
\begin{tabular}{l|l}
Clifford & Gate primitives\\
\hline
\hline
1 & I\\
2 & X/2 X/2\\
3 & Y/2 Y/2\\
4 & Z\\
5 & X/2 -Z/2\\
6 & X/2 Z/2\\
7 & -X/2 Z/2\\
8 & -X/2 -Z/2\\
9 & Y/2 Z/2\\
10 & Y/2 -Z/2\\
11 & -Y/2 -Z/2\\
12 & -Y/2 Z/2\\
13 & X/2\\
14 & -X/2\\
15 & Y/2\\
16 & -Y/2\\
17 & Z/2\\
18 & -Z/2\\
19 & Z Y/2\\
20 & Z -Y/2\\
21 & -X/2 Z\\
22 & X/2 Z\\
23 & X/2 X/2 -Z/2\\
24 & -Z/2 X/2 X/2
\end{tabular}
\begin{caption}{
Gate primitives that comprise each Clifford.}
\end{caption}
\label{table:RB}
\end{table}

The average Clifford time is given by $T_{avg} = (24(t_g+t_c) + 9(2\pi/A_{CDD}))/24 = 92.7 \text{ ns}$ where $t_g+t_c$ is the duration of $\pm X/2$ and $\pm Y/2$ gates. The decomposition of $\pm X/2$, $\pm Y/2$ and $Z$ gates into Cliffords is provided in Table~\ref{table:RB}.

\section{Experimental setup}

\begin{figure}
    \centering
    \includegraphics[width = 3.3in]{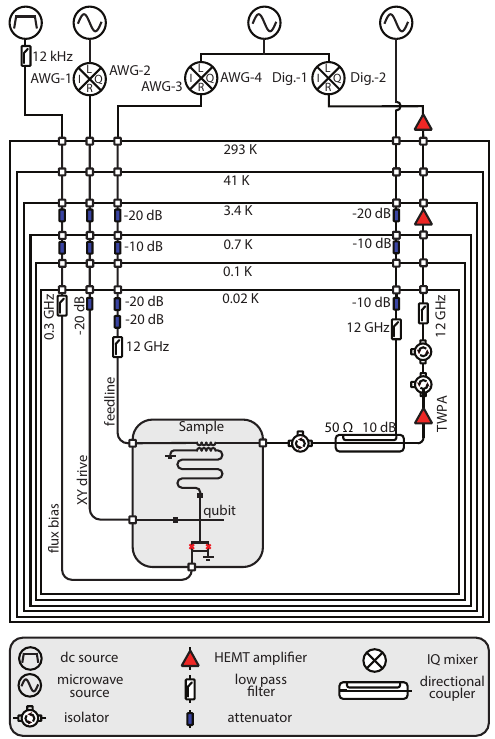}
    \caption{Standard transmon control setup used for CDPQ demonstration.}
    \label{fig:circuit}
\end{figure}

The experimental setup for a CDPQ is identical to that used for controlling a bare transmon superconducting circuit as shown in Fig.~\ref{fig:circuit}. We used a Yokogawa GS200 dc source for flux biasing the qubit. Filtering the Yokogawa source with a $12$~kHz LC low pass filter at room temperature improves the qubit coherence by a factor of ten when the qubit is biased away from its sweet spot. The microwave sources are Rohde-Schwarz SGS100A with all the options. We used Keysight M3202A 1 GS/s arbitrary waveform generators (AWGs) for pulse and CDD generation. 

\bibliographystyle{apsrev4-2}
\bibliography{Main}
\end{document}